\documentclass[sigconf]{acmart}

\usepackage{soul}
\renewcommand\footnotetextcopyrightpermission[1]{} 
\AtBeginDocument{%
  \providecommand\BibTeX{{%
    \normalfont B\kern-0.5em{\scshape i\kern-0.25em b}\kern-0.8em\TeX}}}

\setcopyright{acmcopyright}
\copyrightyear{2023}
\acmYear{2023}
\acmDOI{XXXXXXX.XXXXXXX}

\usepackage{multirow}
\usepackage{tabularray}

\begin{document}
\pagestyle{plain}
\title[]{Investigating impact of bit-flip errors in control electronics on quantum computation}


\author{Subrata Das}
\email{sjd6366@psu.edu}
\affiliation{%
  \institution{The Pennsylvania State University}
  \city{State College}
  \state{Pennsylvania}
  \country{USA}
  \postcode{16801}
}

\author{Avimita Chatterjee}
\email{amc8313@psu.edu}
\affiliation{%
  \institution{The Pennsylvania State University}
  \city{State College}
  \state{Pennsylvania}
  \country{USA}
  \postcode{16801}
}

\author{Swaroop Ghosh}
\email{szg212@psu.edu}
\affiliation{%
  \institution{The Pennsylvania State University}
  \city{State College}
  \state{Pennsylvania}
  \country{USA}
}


\begin{abstract}

In this paper, we investigate the impact of bit flip errors in FPGA memories in control electronics on quantum computing systems. FPGA memories are integral in storing the amplitude and phase information pulse envelopes, which are essential for generating quantum gate pulses. However, these memories can incur faults due to physical and environmental stressors such as electromagnetic interference, power fluctuations, and temperature variations and adversarial fault injections, potentially leading to errors in quantum gate operations. To understand how these faults affect quantum computations, we conducted a series of experiments to introduce bit flips into the amplitude (both real and imaginary components) and phase values of quantum pulses using IBM's simulated quantum environments, FakeValencia, FakeManila, and FakeLima. Our findings reveal that bit flips in the exponent and initial mantissa bits of the real amplitude cause substantial deviations in quantum gate operations, with TVD increases as high as $\sim$ 200\%. Interestingly, the remaining bits exhibited natural tolerance to errors. We proposed a 3-bit repetition error correction code, which effectively reduced the TVD increases to below 40\% without incurring any memory overhead. Due to reuse of less significant bits for error correction, the proposed approach introduces maximum of 5-7\% extra TVD in nominal cases. However, this can be avoided by sacrificing memory area for implementing the repetition code. 
\end{abstract}







\maketitle

\section{Introduction}
\label{sec:intro}

Quantum computing represents a revolutionary leap forward in computational capabilities, offering the potential to solve complex problems currently beyond classical computers' reach. Unlike classical computing, which relies on bits as the smallest unit of data, quantum computing uses quantum bits, or qubits, which can exist in multiple states simultaneously \cite{lapierre2021introduction, nielsen2010quantum}. This fundamental difference enables quantum computers to perform large-scale computations more efficiently \cite{reiher2017elucidating, orus2019quantum, schuld2015introduction, gachnang2022quantum, ajagekar2019quantum}.

Despite the transformative potential of quantum computing, it is intrinsically linked to classical computing systems for development and operation \cite{xu2021qubic, maurya2022compaqt}. For instance, classical systems are essential for qubit control and signal processing, which are critical for the effective functioning of quantum computers. Quantum computers based on superconducting qubits operate using microwave pulses. A quantum program involving tens of qubits may require hundreds to thousands of these pulses per program. The control hardware, functioning at room temperature, dispatches these pulses via coaxial cables to the qubits housed at cryogenic conditions. Quantum computers predominantly utilize a brute-force approach to scale their control hardware. This involves duplicating discrete, off-the-shelf components such as Field Programmable Gate Arrays (FPGAs), Digital to Analog Converters (DACs), and Analog to Digital Converters (ADCs) in line with the number of qubits in the processor. For instance, the control system for the Sycamore chip \cite{arute2019quantum}, which manages 53 qubits, incorporates over 200 DACs, 9 ADCs, and more than 30 FPGAs for effective control and readout. Radio Frequency System-on-Chip (RFSoC) is a type of integrated circuit that combines RF (radio frequency) functions with digital processing capabilities on a single chip. This integration facilitates the development of more compact, efficient, and high-performance devices for applications that require significant RF processing alongside substantial computational tasks. RFSoCs combine an FPGA, a processor, DACs, and ADCs into a single chip, commonly employed in telecommunications applications. These systems are capable of synthesizing signals with bandwidths reaching tens of gigahertz. The recently introduced `QICK' \cite{sussman2022qick} and `ICARUS-Q' \cite{park2021arxiv} platforms utilize RFSoCs for qubit control, showcasing their applicability in advanced quantum computing tasks.

\subsection{Importance of Control Electronics in Quantum Computing}

Quantum circuits are crucial for manipulating information in quantum computing. These circuits consist of quantum gates arranged in a sequence to perform complex computations through the coherent manipulation of qubits. In a typical superconducting quantum computer, quantum gates correspond to microwave pulses that act upon qubits.

\begin{figure*}
    \centering
    \includegraphics[width=1\linewidth]{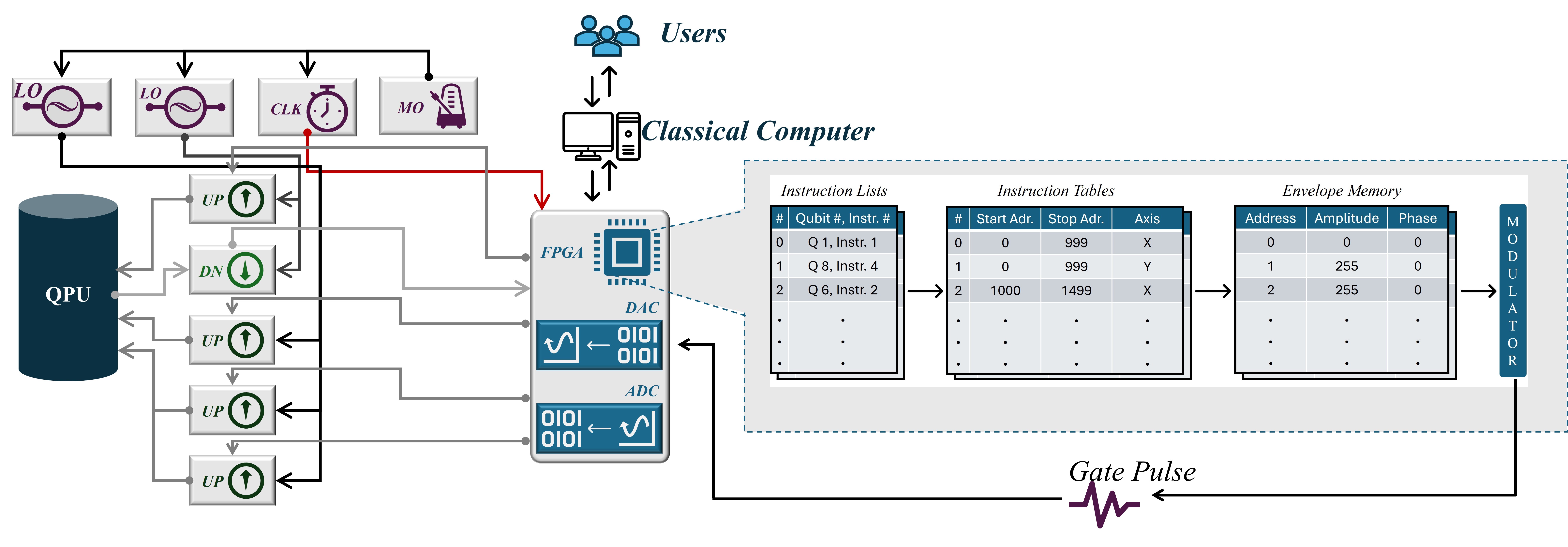}
    \caption{\textbf{Qubit control system hardware and instruction Cycle in quantum computing using FPGA:} Key components include the master oscillator (MO), clock (CLK), local oscillator (LO), field-programmable gate array (FPGA), digital-to-analog converter (DAC), analog-to-digital converter (ADC), upconverter (UP) and downconverter (DN). The diagram also depicts the instruction cycle, where the FPGA memory retrieves and processes amplitude and phase details for each quantum gate, converting these details via a modulator into precise RF pulses directed at the quantum processor unit (QPU), enabling effective quantum operations.
    }
    \label{fig:control_instruction}
\end{figure*}

A comprehensive depiction of a qubit control system is presented in Fig. \ref{fig:control_instruction} integrating both the architectural and operational aspects of FPGA-based RF control systems used in quantum computing. This system, built on advanced FPGA technology, provides a robust platform for qubit pulse generation and quantum state measurement, enabling fully parametric waveform generation that offers complete control over all operational layers \cite{xu2021qubic}. It integrates room temperature electronics hardware, FPGA gateware (i.e., the module consisting of ADC, DAC and FPGA), and engineering software to accurately generate and measure RF pulses for qubit control. Furthermore, this system serves as a platform for developing and refining real-time feedback control strategies, such as rapid reset \cite{vijay2012stabilizing, riste2012feedback}. The hardware utilizes the heterodyne method for compact RF signal generation and detection. It comprises three primary modules: an FPGA/ADC/DAC module for producing and detecting the intermediate frequency (IF) signal, an RF mixing module to shift the signal frequency to or from the desired frequency, and a local oscillator (LO) generation module that delivers low-noise LO signals essential for accurate qubit manipulation.

The process of implementing a quantum circuit using a Quantum Processing Unit (QPU) begins with the user specifying the desired quantum circuit, as illustrated in Fig. \ref{fig:control_instruction}. 
The user submits the specifications of the desired quantum circuit to a cloud interface, where it undergoes preliminary processing and optimization for hardware compatibility. The refined instructions are then relayed to the FPGA, the central controller, which directly manages the quantum hardware operations. Inside the FPGA, the instruction cycle commences with the retrieval of the corresponding pulse for each specified quantum gate from the FPGA’s memory. This memory is meticulously organized to store detailed amplitude and phase information essential for generating the precise pulse for each gate operation. Specifically, if the quantum system is designed to handle five qubits and is capable of implementing five different gates, the FPGA's memory will contain a dataset of 25 unique pulses, effectively covering every possible combination of qubit and gate. Once the appropriate pulse data is identified, it is extracted from the memory. This data includes specific values for amplitude and phase that are crucial for the accurate execution of the quantum gates. Following retrieval, this pulse information is processed through the DAC, which converts the digital pulse data into an analog signal. This signal is then modulated and amplified to ensure it meets the exact requirements for interacting with the qubits. The final output from the modulator, now an RF pulse, is fed into the QPU. Concurrently, the ADC monitors the outputs from the QPU, capturing the results of quantum operations. These results are then processed by the FPGA and sent back through the cloud interface to the user, providing data for real-time adjustments and further analysis.

\begin{table*}[ht]
\centering
\caption{Sources of Faults in FPGA Memory}
\label{tab:faults_fpga}
\begin{tabular}{|p{0.20\textwidth}|p{0.35\textwidth}|p{0.35\textwidth}|}
\hline
\textbf{Mode of Fault} & \textbf{Fault Origins} & \textbf{Impact on FPGA Memory} \\
\hline
Electromagnetic Interference (EMI) & 
- External sources like nearby electronic devices, power lines \cite{rollins2010comparison}. & 
- Induces errors in electronic components controlling qubits. \\
\hline
Power Fluctuations & 
- Unstable power supply. \newline 
- Supply noise \cite{salami2018fault}. & 
- Can lead to bit flips, disrupt operations.\\
\hline
High-Precision Timing Requirements & 
-  Inaccurate timing in qubit manipulation \cite{carmichael2000correcting}. & 
- Can cause memory faults like bit flips. \\
\hline
Aging of Hardware & 
- Bias temperature instability \cite{wirthlin2003reliability}.\newline 
- Hot carrier injection, and electromigration. & 
- Leads to memory faults like bit flips over time. \\
\hline
Manufacturing Defects & 
- Imperfections in the manufacturing process \cite{microsemi2017single}. \newline
- Weak shorts or opens during manufacturing. & 
- Might not be evident initially but can manifest as frequent bit flips occur under certain conditions. \\
\hline
Adversarial Fault Injection & 
- Non-invasive methods such as elevating ambient temperatures, applying radiation. \newline 
- Overloading FPGA with intense computational demands \cite{rakin2021deep}.\newline 
- Creating instability through power supply noise. & 
- Allows adversaries to compromise the system without physical interaction. \newline 
- Specifically possible in shared environments, where quantum computing resources are utilized by multiple parties, enabling indirect interference with critical computing operations. \\
\hline
\end{tabular}
\end{table*}

In quantum computing, where precise control is paramount, FPGAs leverage Block RAM (BRAM) for tasks like storing pulse envelopes used in qubit gate calibration. BRAM is particularly well-suited to this role due to its ability to efficiently store real and imaginary parts of the complex numbers and facilitate rapid, frequent access to these data blocks, crucial for maintaining high-speed operations. It offers significant advantages including the efficient organization and retrieval of complex data, notably the in-phase and quadrature-phase components of control pulses.
Moreover, BRAM's fast access speeds outperform those of distributed or external memory solutions, crucial for operations where timing and synchronization are critical. Additionally, its flexibility to be dynamically configured for various data types and its capability for efficient repeated access to stored data allow FPGAs to quickly adapt and reuse pulse envelopes, even as operational parameters like carrier frequency or phase vary. This makes BRAM an invaluable resource in the architecture of quantum computing systems, enhancing both performance and efficiency.

\subsection{Understanding Bit Flip Errors in Classical Memory}

FPGAs, being classical devices, are susceptible to bit-flip errors. Bit flip errors in the BRAMs of an FPGA can occur due to several factors, often related to the physical and environmental conditions in which the FPGA operates such as radiation effects, electrical noise, and temperature variations \cite{salami2018fault, rakin2021deep, rollins2010comparison}. The BRAMs in FPGAs used in quantum computer control electronics undergo further challenges like intense operations to meet the high performance requirements which can create additional noise in the system (e.g., power supply droop). This, in turn, can cause read/write/retention failures. Such intense operations and subsequent errors can also be adversary induced. For example, adversary sharing the same quantum hardware as victim in multi-tenant computing environment can write a Trojan program to generate frequent pulses increasing power consumption and subsequent supply noise. An insider adversary can also consider non-invasive tampering of the FPGA to inject bit flip errors by manipulating ambient conditions. The impact of bit flip errors in BRAMs can be significant due to the precision and sensitivity required in quantum computing. Some specific reasons that could lead to bit flip errors or similar memory disturbances in quantum computer control electronics are tabulated in Table \ref{tab:faults_fpga}. To make the matter worse, current state-of-the-art FPGAs do not support ECC (Error Correcting Code) on BRAMs by default. For example, single port BRAM in Xilinx FPGAs are not ECC protected \cite{XilinxBRAMECC2020}. 

\subsection{Translating Classical Errors to Quantum Errors}

\begin{figure}[t]
    \centering
    \includegraphics[width=1\linewidth]{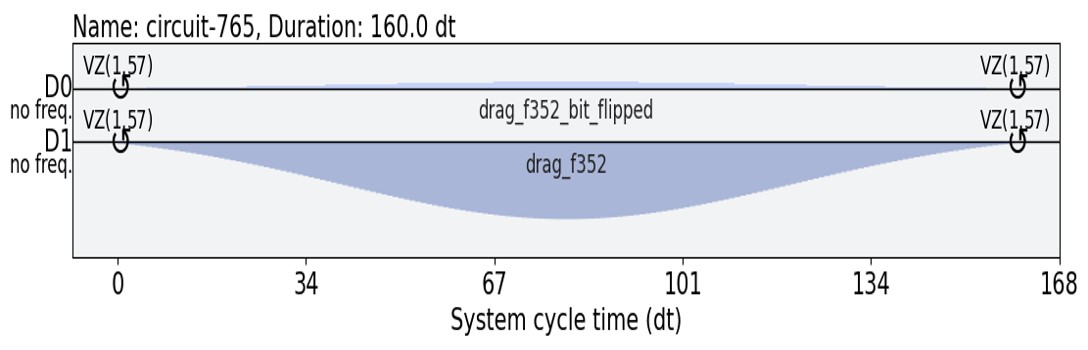}
    \caption{Pulse shape impact from bit flip in Hadamard Gate. The bottom graph shows the original pulse amplitude for IBM's FakeValencia Hadamard Gate, and the top graph illustrates the drastic reduction in amplitude caused by a third-bit flip in FPGA memory.}
    \label{fig:pulse}
\end{figure}

Bit flip errors in FPGA memory can disrupt quantum operations by causing faulty pulse generation. Fig. \ref{fig:pulse} illustrates the pulse shape of a Hadamard Gate from IBM's FakeValencia Backend obtained using Qiskit. The bottom part of the figure shows the actual pulse shape, with the amplitude represented as a 32-bit floating-point number stored in the FPGA memory. If a bit flip occurs, such as a flip in the third bit, the amplitude undergoes a dramatic decrease from its original value of 0.09618851775276127+0.0008448724348311288j to a bit-flipped value of 2.239563395844968e-11+0.0008448724348311288j. This significant reduction (also shown in Fig. \ref{fig:pulse}) can lead to substantial errors in the quantum computing outcomes. 

\subsection{Our Contributions}
\label{subsec:contributions}

This paper provides a comprehensive analysis of the effects of bit flip errors in FPGA memory, specifically within the context of quantum computing control systems. 
We have conducted a series of controlled sensitivity analysis where single bits in the amplitude (both real and imaginary components) and phase values of quantum pulses were flipped. These experiments were carried out using various IBM quantum fake backends, which simulate real quantum computing environments. This approach allows for a precise understanding of how such errors affect quantum gate operations. Secondly, our analysis goes beyond general fault impacts by examining the specific effects of bit flips on the functionality and performance of various native quantum gates such as Hadamard, NOT, and Rotation-gates. Finally, we propose repetition code to protect the important bits of the BRAM and mitigate the impact of bit-flip errors.

\subsection{Paper Organization}
\label{subsec:paper_organization}

In the remainder of this paper, Section \ref{sec:background} provides a background on the role of FPGAs in quantum control systems and explains the basics of quantum pulse specification and data representation. Section \ref{sec:analysis} details our experimental setup and findings on the bit flip errors. Section \ref{sec:strategies} presents the analysis with 3-bit repetition code to mitigate the impact of bit flip errors. Finally, Section \ref{sec:conclusion} draws conclusions.

\section{Background}
\label{sec:background}
\subsection{FPGA in Quantum Control Systems}

FPGAs are integral to the architecture of modern quantum control systems due to their versatility, reconfigurability and performance. 
Qubits exhibit quantum mechanical properties that is harnessed in quantum computing to perform complex calculations more efficiently than classical computers. The role of FPGAs in these systems is multifaceted. They generate precise and high-frequency signals needed to manipulate qubits through quantum gates. This manipulation is critical for executing quantum algorithms. FPGAs can dynamically adjust these signals in response to real-time feedback from the quantum system, which is crucial for maintaining the delicate state of qubits during computations. Moreover, the reconfigurability of FPGAs offers a significant advantage. As quantum algorithms evolve and new quantum error correction codes are developed, the underlying hardware must also adapt. FPGAs allow for these rapid changes without the need for replacing physical components, unlike application-specific integrated circuits (ASICs) which are hard-coded to perform specific tasks.
FPGAs also contribute to error management in quantum systems. They process readout signals from qubits to detect and correct errors on the fly. This capability is vital for quantum error correction, which is essential to counteract the inherent instability of qubits and ensure the accuracy of quantum computations.

The integration of FPGAs into quantum computing extends beyond mere functionality. They enhance the scalability of quantum computers. As systems expand to include more qubits, the complexity of control tasks increases exponentially. FPGAs handle this scalability efficiently, managing multiple tasks simultaneously and sustaining the synchronization required for complex quantum operations.

\subsection{Quantum Pulse Specification and Data Representation in FPGAs}

\subsubsection{Complex Amplitude Components}

In quantum computing, the precise manipulation of qubits is achieved through the application of controlled quantum gate-specific pulses that are complex waveforms. 
Each pulse is characterized by parameters such as amplitude, phase, and duration, where the amplitude is often represented as a complex number with real (I) and imaginary (Q) components. 


The exact shape, timing, and fidelity of the pulse can determine the success of quantum gate operations. The real (I) and imaginary (Q) parts of the pulse amplitude are used to finely control the qubit's trajectory through its state space, which is fundamental in executing coherent quantum mechanical manipulations. This I/Q representation is analogous to the I/Q modulation widely used in digital communication for its spectral efficiency. I/Q Modulation involves two carriers that are orthogonal to each other. One is known as the in-phase (I) component, and the other is the quadrature (Q) component. The I component aligns with the cosine function, while the Q component aligns with the sine function, effectively phase-shifted by 90 degrees from the I component. 
In the context of quantum computing pulse sequences, the real part of the amplitude corresponds to the in-phase component, and the imaginary part corresponds to the quadrature component. The real part (I) defines the basic strength and phase of the pulse, while the imaginary part (Q) introduces an additional phase shift that is crucial for manipulating the quantum state of the qubit in a precise manner.

\subsubsection{Phase Control}

Phase control is a critical operation in quantum computing, and its implementation can vary across different platforms. In the context of IBM Quantum systems that utilize Qiskit Pulse, the \texttt{ShiftPhase} instruction provides a refined mechanism for this purpose. This instruction is important as it updates the modulation phase of subsequent pulses on the same channel, enabling precise adjustments necessary for accurate qubit manipulation.
Specifically, this instruction modifies the phase $\phi$ of the output signal on a channel, impacting all following pulses. The operational mechanics of \texttt{ShiftPhase} can be expressed mathematically as:

\begin{equation}
\text{Re}\left[\exp\left(i2\pi f_j dt + \phi\right)d_j\right].
\end{equation}

This equation represents the real part of the pulse generated on a \texttt{PulseChannel}, where $\phi$ is adjusted by the phase operand of the \texttt{ShiftPhase} instruction. Such adjustments are crucial for precise control over the quantum state rotations around the z-axis, enabling nearly error-free Z-rotations.

\subsubsection{Floating Point Representation in FPGA Memory}

Typically, the real and imaginary parts of pulse amplitude, along with values used in phase control instructions like \texttt{ShiftPhase}, are represented as floating-point numbers. Floating point numbers in FPGA-based quantum control systems are typically stored using the IEEE 754 standard where  a floating-point number is represented through three distinct parts:
(a) \textbf{Sign bit:} A single bit that indicates if the number is positive or negative. (b) \textbf{Exponent:} An 8-bit field (for single precision) that scales the number by powers of two. (c) \textbf{Mantissa (or significand):} A 23-bit field that contains the significant digits of the number.
For example, the floating-point representation of the number -248.75 in IEEE 754 format is as follows:
\begin{itemize}
    \item The sign bit is 1, indicating a negative number.
    \item The exponent is encoded with a bias of 127; for the actual exponent of 8, it is stored as 135, or \texttt{10000111} in binary.
    \item The mantissa is derived from the non-integer part of the number, stored as \texttt{11110001100000000000000} in binary.
\end{itemize}

This encoding results in the binary representation \texttt{11000011\allowbreak011110001100000000000000}. The FPGA's floating-point unit must accurately generate and manipulate these IEEE 754 representations to ensure the precision of quantum operations. This capability is particularly important because even minor inaccuracies in pulse specification can lead to significant errors in quantum computations, affecting the reliability and performance of the entire quantum computing system.

\section{Analysis of Bit Flip Errors}
\label{sec:analysis}
\subsection{Experimental Setup}

\begin{figure*}[ht]
    \centering
    \includegraphics[width=1\textwidth]{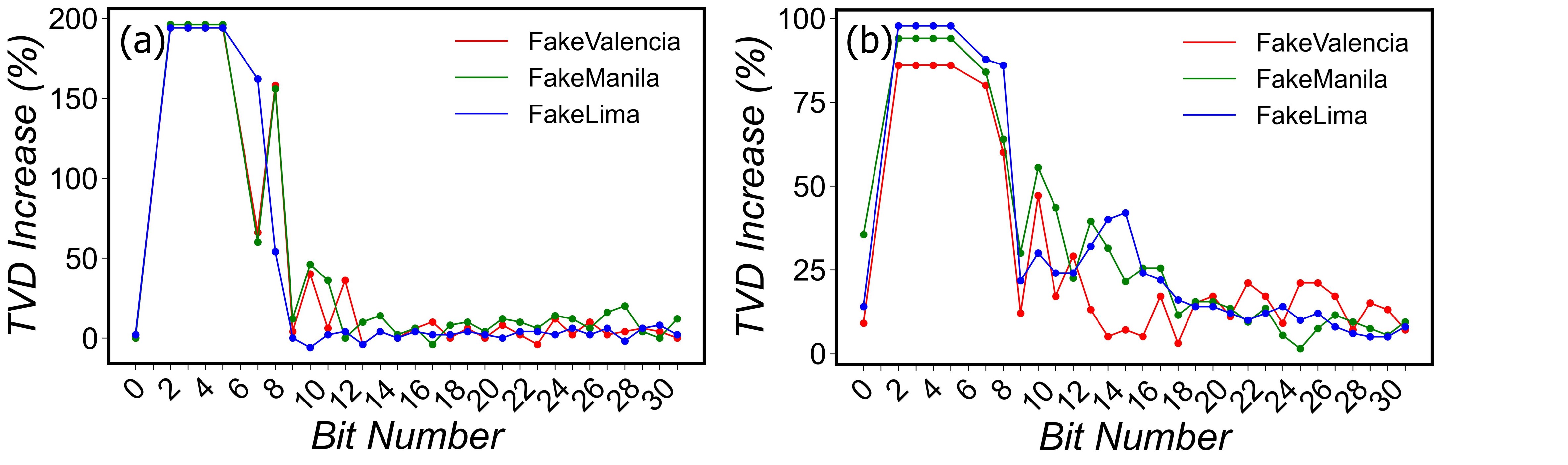}
    \caption{TVD increase from bit flips in the real part of amplitude for (a) X and, (b) H gates across three simulated backends: FakeValencia, FakeManila, and FakeLima demonstrating consistent sensitivity trend across backends.}
    \label{fig:multibackends}
\end{figure*}

We utilized the Qiskit framework to simulate quantum circuits and manipulate quantum pulse sequences on IBM’s fake quantum backends, specifically FakeValencia, FakeManila and FakeLima. These environments allow for the realistic simulation of quantum computational processes without the need for actual quantum hardware.

\subsubsection{Quantum Gate and Pulse Configuration}
We start by constructing a basic quantum circuit for different single and two qubit gates using the QuantumCircuit module. The circuit is then transpiled and scheduled for the specific backend to obtain the corresponding control pulse sequence. This step translates high-level quantum operations into the low-level pulses that physically manipulate the qubits.

Each pulse in the sequence, particularly the Drag pulse associated with the X gate, is analyzed to extract its amplitude and phase components. The Drag pulse, which is designed to minimize leakage to non-computational states, consists of a complex amplitude described by real (I) and imaginary (Q) parts. These components are subsequently converted to a 32-bit floating-point format. Bit flip experiments are performed on this binary representation.

\subsubsection{Bit Flip Simulation and Analysis}
Each bit, from the most significant bit of the mantissa to the sign bit, is flipped at a time to observe the resultant effect on the quantum operation. After flipping a bit, the modified binary string is converted back to a floating-point number, which is then used to adjust the amplitude of the Drag pulse in a custom-built pulse sequence.

\subsubsection{Measurement and Comparison Metrics}
The primary metric used to evaluate the impact of the bit flips is the Total Variation Distance (TVD) which is a statistical measure used to quantify the difference between two probability distributions. In our experiments, TVD is calculated between the probability distributions of the quantum state measurements obtained from the ideal (unmodified) and the perturbed (bit-flipped) pulse sequences. TVD provides a clear quantitative measure of how significantly a single bit flip can alter the behavior of a quantum gate. The reason for using TVD over other metrics is its sensitivity to both small and large discrepancies in distributions, which is critical in quantum computations where even minor deviations can lead to significant errors.
Each experiment is run multiple times to ensure statistical relevance.
This setup not only highlights the specific vulnerabilities of quantum systems to hardware imperfections but also serves as a testbed for evaluating potential error correction techniques that could be implemented at the hardware level.

\subsection{Results and Discussions}


Our experimental investigation began with a systematic study of bit flip errors across multiple fake backends provided by IBM: FakeValencia, FakeManila, and FakeLima. We focused on manipulating the real part of the amplitude of the pulse envelope for quantum gates pulses, specifically the X and H gates, which are fundamental to a wide array of quantum algorithms. For each gate, we flipped each of the 32 bits one at a time in the real component of the amplitude and observed the outcomes. It should be noted that during our experiments, specific exponent bits, notably bit 1 and bit 6, resulted in invalid pulses because the maximum pulse amplitude norm exceeded 1.0. We assume that such extreme conditions would be immediately recognized by the system's error detection protocols to halt the transmission of the pulse. In our experiments, to provide an uninterrupted visual trend, we employed a linear interpolation method to estimate the missing TVD values by connecting the adjacent data points directly.

Interestingly, the trends observed in the impact of bit flips were consistent across the three backends. This uniformity was evident in both the X and H gates, as illustrated in Fig. \ref{fig:multibackends} (a) and (b), respectively. TVD increase measurements showed similar patterns of sensitivity and resilience to specific bit flips, regardless of the backend.

For TVD increase calculations, firstly, we compute the TVD between the outcomes from the bit-flipped pulse settings and the ideal theoretical outcomes, which we refer to as `TVD flipped vs. ideal'. Secondly, we compare the original, unmodified pulse settings against the ideal outcomes, termed `TVD original vs. ideal'. The TVD increase is then quantified as the difference between these two values, effectively measuring the additional deviation caused by flipping a bit in the pulse's amplitude. Notably, flipping the first bit, which is the sign bit, did not show a significant impact on the TVD. This indicates that the changes in the sign of the pulse's amplitude have a minimal effect on the operational characteristics of the gate. Further, detailed observation of bits 1 to 8, which represent the exponent part of the floating-point representation, revealed a consistently high TVD increase. Flipping any of these bits causes substantial changes in the amplitude's magnitude, leading to dramatic shifts in quantum gate behavior. Specifically, bits 9 to 17 within the mantissa also showed significant TVD increases, suggesting these higher-order bits play a crucial role in defining the precision of the pulse amplitude. In contrast, bits 18 to 31, which represent the lower-order part of the mantissa, exhibited a lesser impact on TVD, indicating their relatively minor role in affecting the quantum gate operations across the tested backends. This uniformity in response to bit flips was evident in both the X and H gates, as illustrated in Fig. \ref{fig:multibackends} (a) and (b), respectively. TVD measurements showed similar patterns of sensitivity and resilience to specific bit flips, regardless of the backend.

The consistency in the experimental results across different backends can be attributed to the standardized simulation protocols used in Qiskit, which are designed to mimic the physical behaviors of actual quantum devices. These protocols ensure that the fundamental physics governing quantum operations are uniformly modeled, regardless of the specific backend. Therefore, the effects of modifying bit values in the amplitude's real part are similarly reflected across all simulated environments. Given the similar trends observed and to streamline our experimental process, we chose to continue detailed experiments solely on FakeValencia. This approach allows for more efficient use of resources and simplifies our analysis without sacrificing the breadth of applicability of our results. The choice of FakeValencia, while arbitrary among the three, provides a representative sample that is likely indicative of general behaviors in other simulated quantum computing environments.

\begin{figure}[t]
    \centering
    \includegraphics[width=0.45\textwidth]{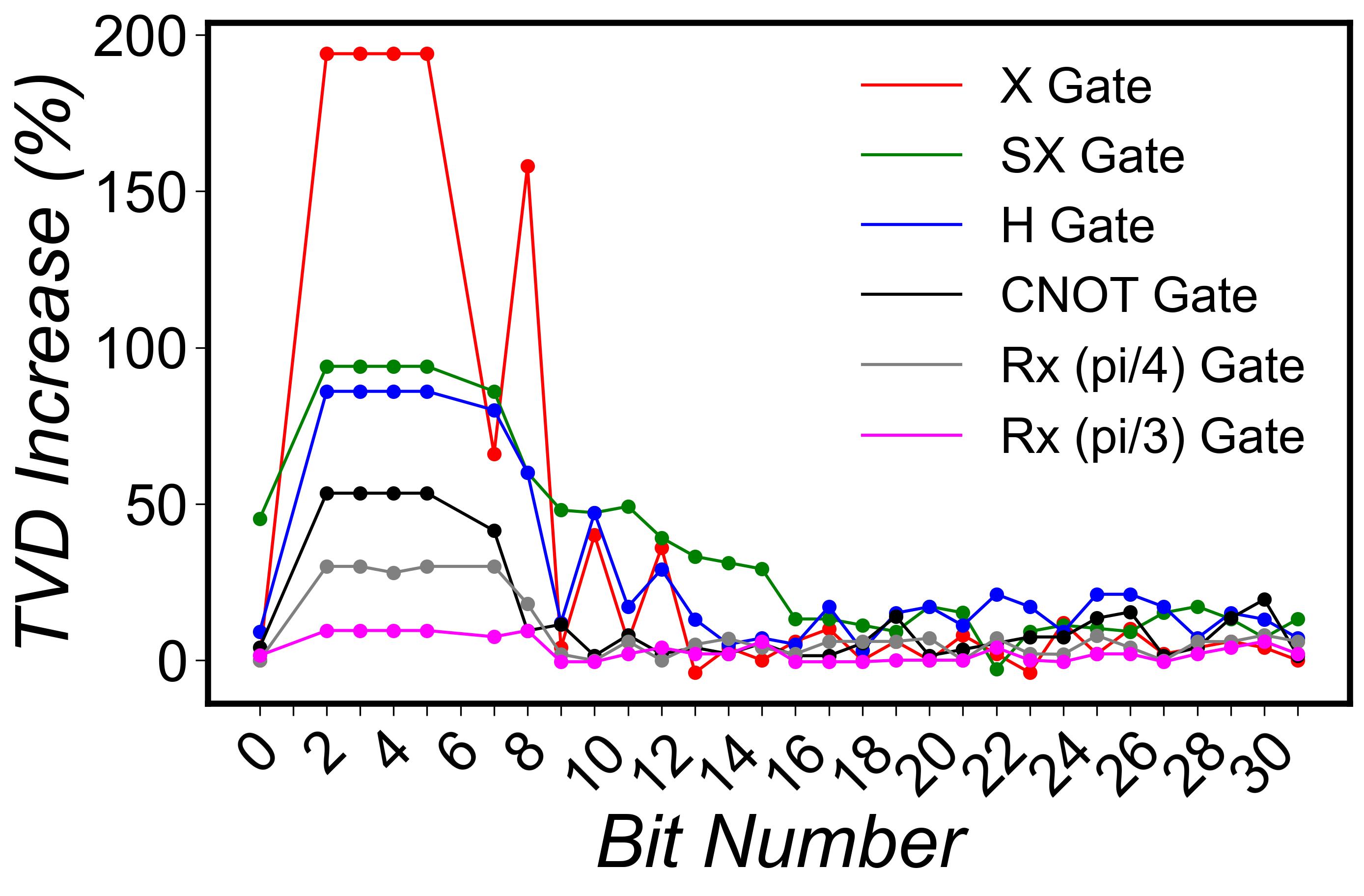}
    \caption{TVD increase from bit flips in the real part of amplitude for various quantum gates simulated in FakeValencia backend.}
    \label{fig:gates_real}
\end{figure}

We extended our experiments to include additional quantum gates such as the SX, CNOT, and rotation gate, Rx at $\pi/4$
 and $\pi/3$ angles. These experiments focused on how bit flips in the real part of the pulse amplitude affect the functionality of these gates, as shown in Fig \ref{fig:gates_real}. 
We note that the exponent bits (bits 1-8) of the floating-point representation have a significant impact on the amplitude, greatly affecting gate operations when altered. For example, flipping just one exponent bit in the X gate led to a nearly 200\% increase in TVD. This happens because the exponent bits scale the amplitude exponentially, leading to large changes in the pulse's magnitude. Such changes can result in incorrect quantum operations, highlighting the importance of these bits for the accuracy of quantum gates. The SX and H gates also show a strong sensitivity to changes in the exponent bits, but to a lesser extent than the X gate.

In contrast, the Rx and CNOT gates demonstrated more resilience to these bit flips. This resilience could be due to the specific roles these gates play in quantum circuits. The Rx gates, which perform rotations around the x-axis by specific angles, might be less affected by small amplitude changes because their primary function is to change the phase of the qubits rather than their state. As for the CNOT gate, its operation as a two-qubit control gate might inherently buffer it against the impact of amplitude changes in single-qubit control lines. Its functionality relies more on the relative states of two qubits rather than on the precise amplitude of a control pulse.

Additionally, bit flips beyond bit 17 showed minimal effect across all gates tested. This is consistent with the nature of the mantissa in floating-point numbers, where the lower-order bits influence only the precision of the pulse amplitude. These details have limited impact on the quantum gates' operational characteristics, demonstrating the gates' ability to perform correctly despite minor perturbations. 
\begin{figure}[t]
    \centering
    \includegraphics[width=0.45\textwidth]{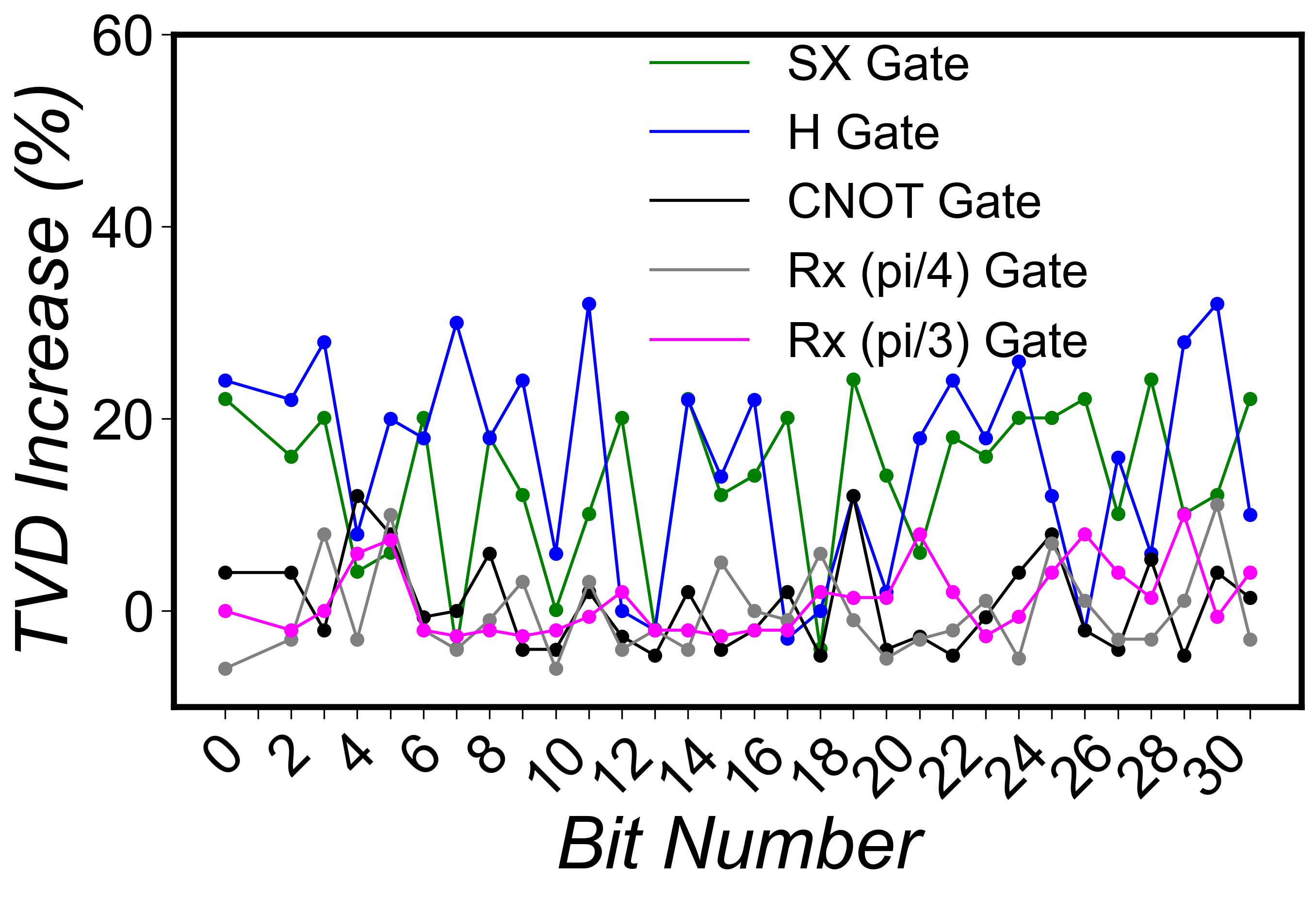}
    \caption{TVD increase from bit flips in the imaginary part of amplitude for various quantum gates simulated in FakeValencia backend.}
    \label{fig:gates_imaginary}
\end{figure}

We also analyzed the effects of bit flips in the imaginary part of the pulse amplitude across various gates, except the X gate which does not utilize an imaginary component in its amplitude. 
From Fig. \ref{fig:gates_imaginary}, it is clear that the Total Variation Distance (TVD) increase shows no consistent pattern or trend across the different bit positions, whether they are in the exponent or the mantissa. 
This diminished sensitivity can be attributed to the relatively smaller magnitude of the imaginary part of the pulse amplitude compared to its real counterpart. For example, the amplitude for the H gate on FakeValencia backend is recorded as  0.09618851775276127+0.0008448724348311288j, where the imaginary part is much smaller than the real part. Therefore, alterations to the imaginary component through bit flipping have a reduced effect on the gate's functionality due to its lower numerical influence.

Additionally, similar to bit flips in the real part, the Rx and CNOT gates demonstrate greater resilience to errors in the imaginary part. This resilience could stem from the structural and operational characteristics of these gates. The Rx gates, designed to perform precise rotational operations, and the CNOT gate, functioning as a two-qubit control gate, may inherently possess robust error handling mechanisms that mitigate the impact of slight amplitude variations, particularly those affecting the less significant imaginary part.

\begin{figure}[t]
    \centering
    \includegraphics[width=0.45\textwidth]{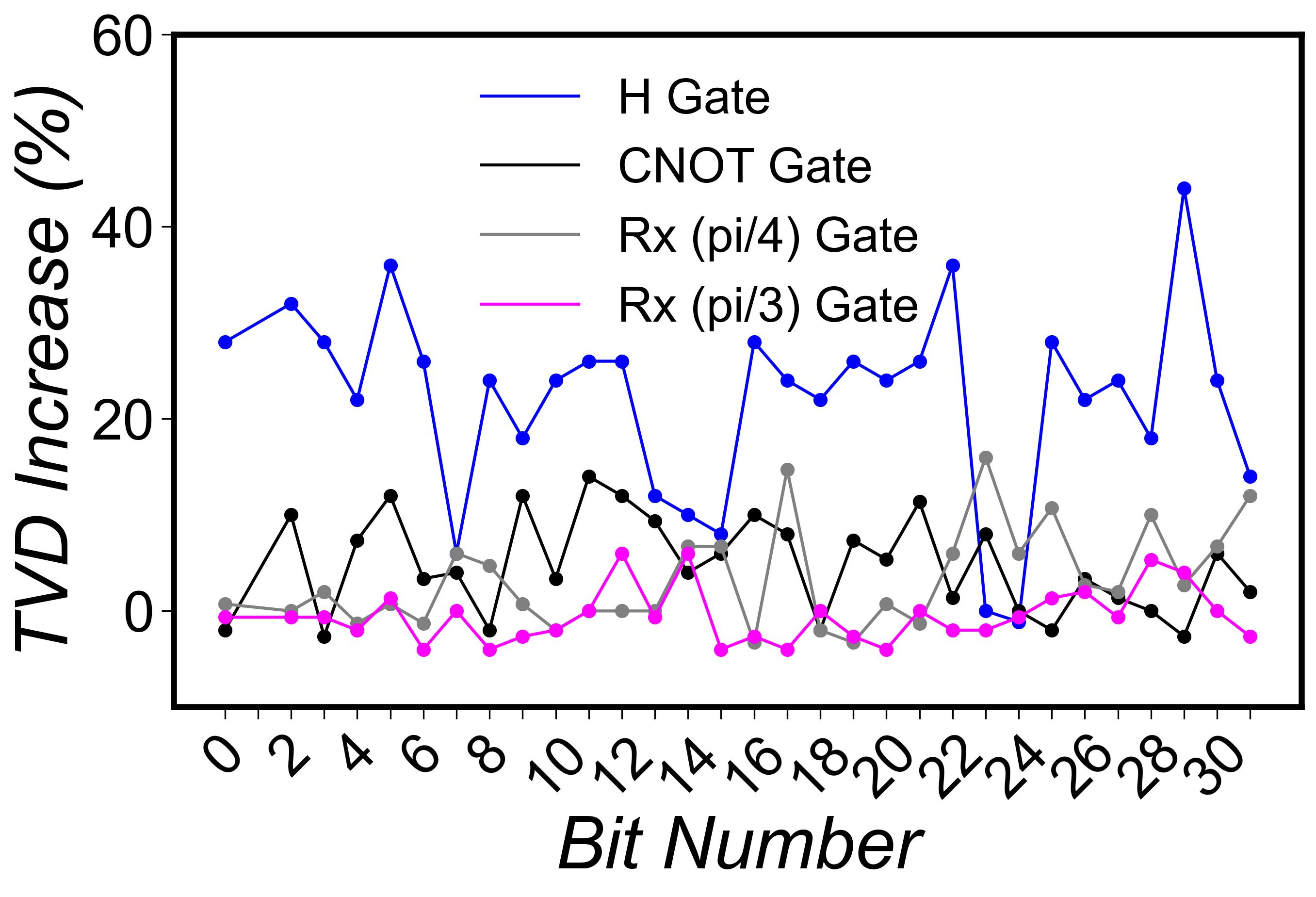}
    \caption{TVD increase from bit flips in the phase for various quantum gates simulated in FakeValencia backend.}
    \label{fig:gates_phase}
\end{figure}

In the next experiment, we analyzed the impact of bit flips on the phase shift values of quantum gates utilizing the \texttt{ShiftPhase} instruction. The results, shown in Fig. \ref{fig:gates_phase}, revealed that bit flips in phase shift values do not follow a consistent pattern across different bits, either in the exponent or mantissa. This observation suggests a uniform sensitivity to phase errors across the binary representation of the phase values, which is crucial for understanding the robustness of quantum gates against such errors. The TVD impacts were generally lower compared to those from amplitude bit flips, indicating a milder impact on quantum operations. This reduced sensitivity can be understood in terms of the specific role of the \texttt{ShiftPhase} instruction, which is critical as it updates the modulation phase of subsequent pulses on the same channel. This instruction does not alter the pulse amplitude but adjusts the phase angle $\phi$ of the output signal, impacting all following pulses. Such phase adjustments are essential for the precise control required in quantum operations, and the fact that they do not directly change the amplitude means that errors in phase values may have a less disruptive impact on the overall quantum state compared to amplitude errors. Gates such as the Rx and CNOT showed greater resilience to these phase errors, likely due to their inherent error handling capabilities within their operational structures.

\section{Mitigating Bit Flip Error}
\label{sec:strategies}
From analysis, we observed that flips in the real part specifically in the exponent bits significantly disrupted operations, unlike the flips in the imaginary and phase parts.
Given the greater impact of real part errors, we propose repetition code to enhance the resilience of quantum computing systems.


\subsection{3-bit Repetition Codes for Error Resilience}

\begin{figure}[t]
    \centering
    \includegraphics[width=0.5\textwidth]{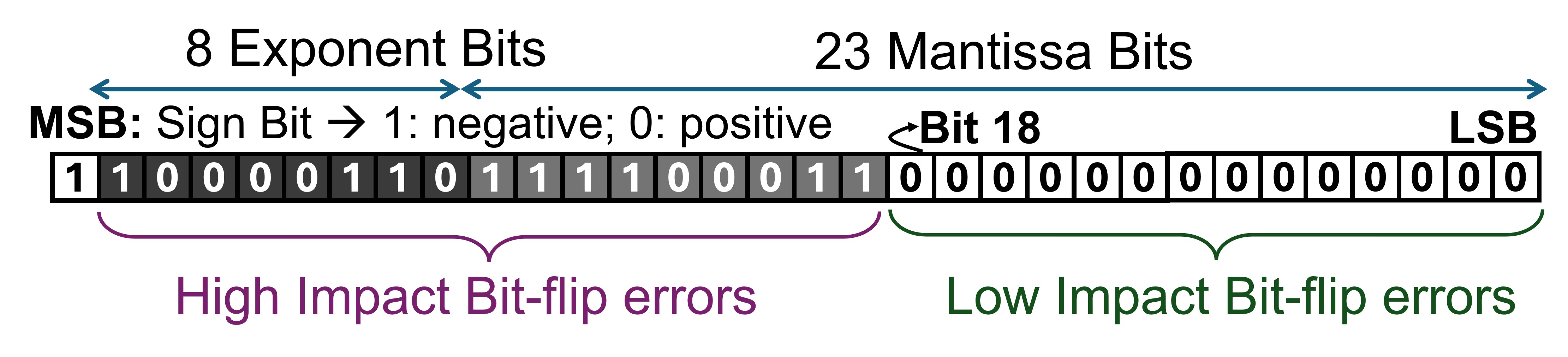}
    \caption{\textbf{Impact of bit flips in the real part of the amplitude of quantum gate pulses as represented in FPGA memory.} A 32-bit IEEE 754 floating point number $-248.75$, as stored in FPGAs for generating quantum computer gate pulses. The format includes a Most Significant Bit (MSB) for the sign, followed by 8 exponent bits and 23 mantissa bits. The bits are categorized into two groups based on experimental results. Darker shaded areas (bit 1 to 17) highlight the regions where bit flips significantly affect the accuracy of the gate pulses, while lighter shaded areas (bit 0, bit 18-31) indicate regions with low or negligible impact. }
    \label{fig:bit_storage_in_FPGA}
\end{figure}

In our study of bit flip effects on the real part amplitude of quantum gate pulses, the most significant impacts were observed in the exponent bits (bits 1-8) and the first nine bits of the mantissa (bits 9-17), as indicated by the darker shading in Fig \ref{fig:bit_storage_in_FPGA}. 
Conversely, bits beyond the 17th position and the sign bit showed minimal effect on operational outcomes, marked by lighter shading in the figure. 
Given that bits 1 and 6 of the exponent consistently resulted in invalid pulses when flipped exceeding the maximum allowable pulse amplitude, we assume the system will detect these errors and halt operations. This assumption allows us to strategically exclude these bits from our error correction efforts. We can then focus on safeguarding the remaining impactful bits (2-5, 7-17).

We propose employing a 3-bit repetition code specifically for the seven bits that most significantly affect quantum gate operations: bits 2, 3, 4, 5, 7, 8, and 9. This error-correcting strategy involves encoding each of these critical bits into three bits. For example, if a bit originally has a value of `1', it will be encoded as `111'; similarly, a  `0' would be encoded as `000'. This tripling of each bit enhances the system's ability to detect and correct single bit errors by allowing a majority vote among the three bits to determine the correct value. 
In majority voting, the most common value among the three replicated bits is chosen as the correct value. For instance, if the received triplet is `010', the system recognizes that an error has occurred and corrects it to `000' based on the majority of bits being `0'.

To implement error correction without increasing the overall memory overhead, we repurpose fourteen less critical bits (specifically, bits 18-31) to encode each of the seven high-impact bits (bits 2, 3, 4, 5, 7, 8, and 9) into three bits. This reallocation effectively utilizes the memory space that would otherwise hold less crucial data. The bits in positions 10-17, while still impactful, will not be encoded due to memory constraints, balancing the need for error correction with the available memory resources.

The effectiveness of this coding scheme is described using the $[n,k,\delta]$ notation, where $n$ is the number of bits in the encoded word, $k$ is the number of original data bits, and $\delta$ is the Hamming distance of the code. For the 3-bit repetition code used here, the parameters are as follows: $n=3$ indicates that each bit is expanded to three bits, $k=1$ shows that each triplet originates from one original bit, and $\delta=3$ represents the Hamming distance. The Hamming distance of three means that the code can detect up to two errors and correct one error within each triplet of bits. 

\subsection{Evaluating Effectiveness of Repetition Code }

\begin{figure}[t]
    \centering
    \includegraphics[width=0.45\textwidth]{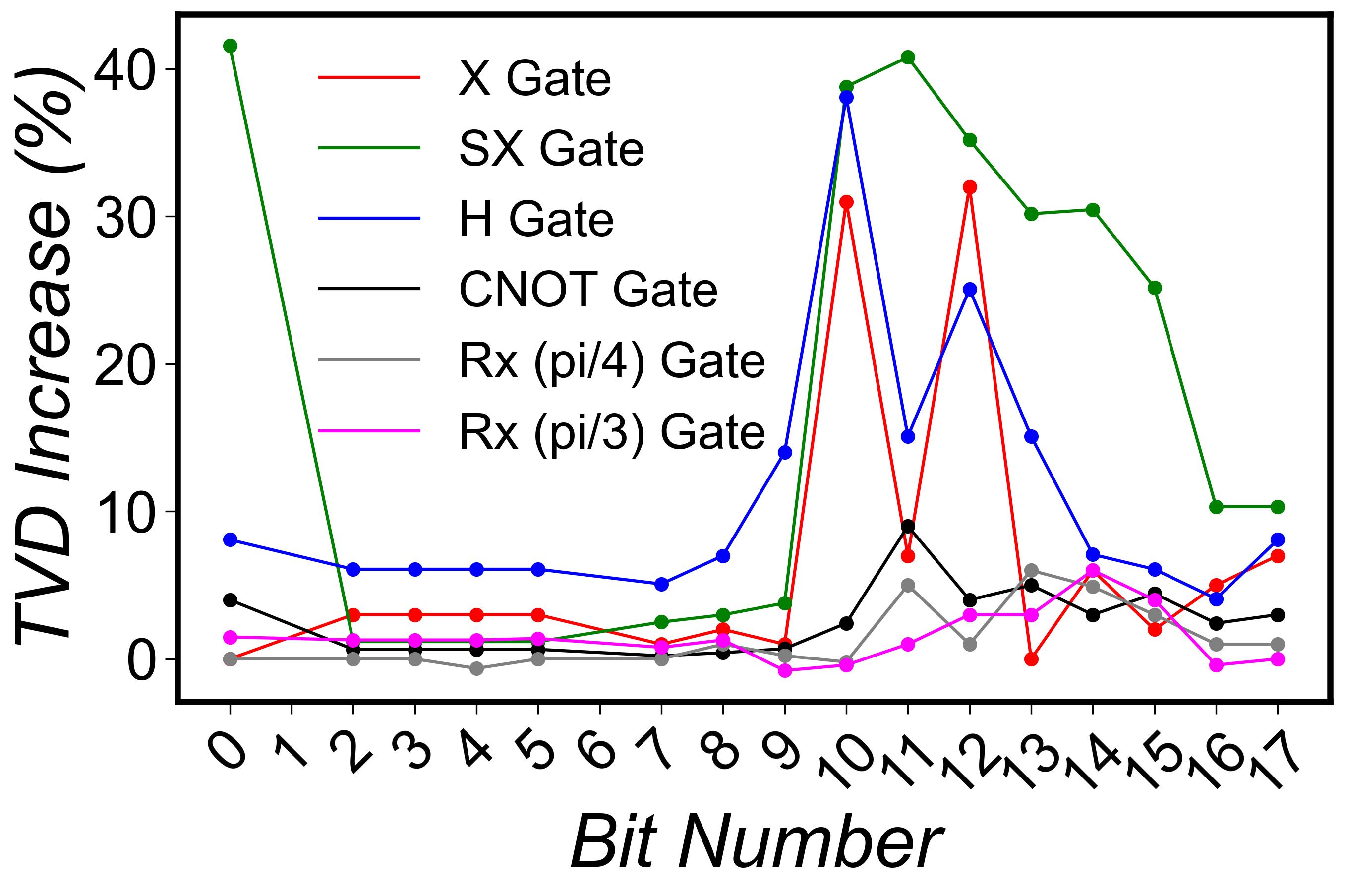}
    \caption{\textbf{Reduction in TVD for quantum gates with 3-bit repetition code on bits 2, 3, 4, 5, 7, 8, and 9.}}
    \label{fig:gates_real_errorcorrection}
\end{figure}

\begin{figure}[t]
    \centering
    \includegraphics[width=0.45\textwidth]{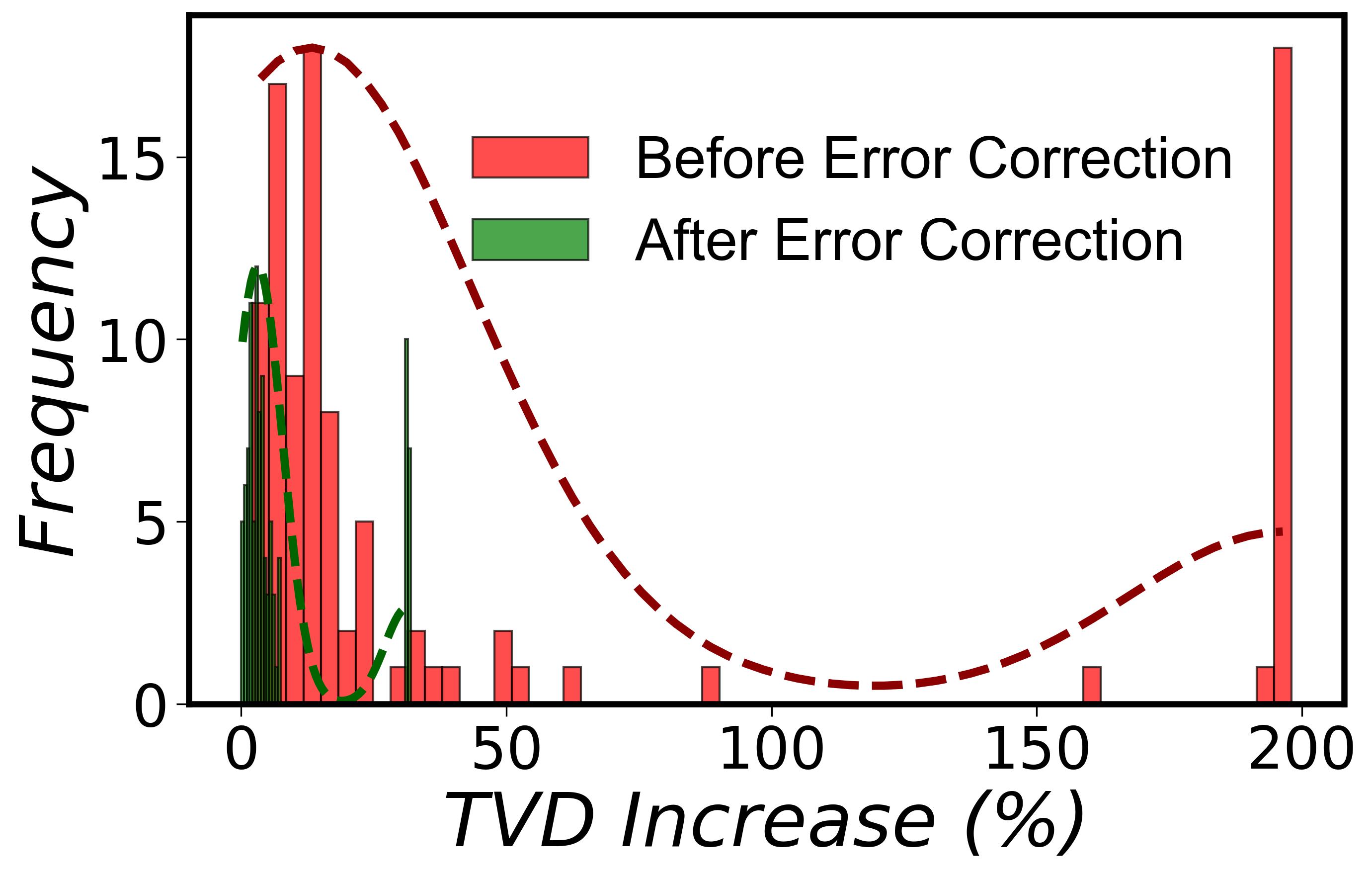}
    \caption{\textbf{TVD increase distribution for X gate pulses with 100 random bit flips.} We show the TVD increases before and after error correction. The narrowed TVD distribution after 3-bit repetition error correction, highlights it’s effectiveness.}
    \label{fig:distribution}
\end{figure}


The results, shown in Fig. \ref{fig:gates_real_errorcorrection}, indicate that TVD significantly decreased for bits 2, 3, 4, 5, 7, 8, 9 after error correction (e.g., below 7\% compared to $\sim$ 200\% before correction). 
When a bit flip is detected within these encoded triplets, the majority voting process corrects it, effectively restoring the original bit value.
However, it is important to note that the TVD did not drop to zero for the error-corrected bits. This can be attributed to the loss of some information due to our strategy of discarding the less impactful bits (18 to 31) to make room for encoding the more critical bits. This trade-off was necessary to balance error correction with memory constraints. For the bits between 10 and 17, which we could not encode due to memory constraints, the TVD was higher compared to the error-corrected bits but still remained under 40\%. This observation highlights that even without applying error correction to all bits, our system still demonstrates strong resilience. Note that implementing the 3-bit repetition code for 7 specific bits involves additional logic circuits to encode each bit into triplets and to handle the majority voting mechanism. These modifications require extra FPGA resources, which could affect the overall system efficiency and complexity. However, the benefits of improved error correction and system reliability often justify the minor added complexity in critical applications like quantum computing.

Finally, we conducted experiments to rigorously test the effectiveness of our error correction strategy under simulated real-world conditions. Over the course of 100 runs, we randomly flipped one bit at a time in the real part of the amplitude of the X gate. This methodology was crucial for evaluating how the system handles spontaneous errors that might occur during typical operations. The initial distribution of TVD increases, shown in Fig. \ref{fig:distribution} (red plot) before implementing the error correction, displayed a wide spread. Frequencies peaked sharply at $\sim$ 13\%, indicating small change for some flips, and soared to nearly 200\% for others, highlighting flips that caused severe disruptions in gate functionality. Upon applying the 3-bit repetition error correction method, the TVD distribution, as illustrated in Fig. \ref{fig:distribution} (green plot), narrowed markedly. Most increases were now tightly clustered around 3\%, with no instances exceeding 35\%. This significant reduction in TVD highlights the robust effectiveness of our error correction strategy. 

\subsection{Other Error Correction Techniques}
One can use SECDED (Single Error Correction, Double Error Detection) \cite{hamming1950error} instead of repetition code for better fault coverage at lower overhead. 
The number of parity bits (\(k\)) needed to protect 24 data bits (in our case) ((\(n\)) can be calculated as follows:
\[2^k - 1 \geq n + k.\]
The smallest \(k\) that satisfies this condition for \(n = 24\) is 5, as \(2^5 - 1 = 31\) meets the requirement \(24 + 5 = 29\). Therefore, 5 parity bits can effectively manage 24 data bits. The 5 bits from the less impactful range of 24-31 can be repurposed to serve as these parity bits. 

By analyzing the TVD data without error correction, we can estimate improvements in TVD after applying SECDED to the first 24 bits of quantum gate real amplitudes. This estimation assumes that the maximum TVD will occur for the remaining bits that are not corrected. For example, we project the TVD for the X gate to decrease to between 10\% and 15\%. Similar reductions are expected for other gates, with the H and SX gates likely seeing TVD decreases to around 20-25\% and 15-20\%, respectively, and the CNOT gate to between 15\% and 20\%. The rotation gates, Rx($\pi/4$) and Rx($\pi/3$), are anticipated to achieve even lower TVDs, ranging from 5\% to 10\% and 5\% to 7\%, respectively. These estimates also take into account the impact of discarding data from the 5 least impactful bits, which could contribute to residual TVD.

\section{Conclusion}
\label{sec:conclusion}

We investigated the impact of bit-flip errors in FPGA memories that store amplitude and phase information used to generate quantum gate pulses. We identified significant disruptions in quantum operations due to faults in the real part of the amplitude, particularly within the exponent and initial mantissa bits. We showed that a 3-bit repetition code can effectively reduce these errors, substantially enhancing the reliability of quantum gate operations. Our analysis also indicated that quantum computing is naturally resilient to errors in general. 

\begin{acks}
The work is supported in parts by the National Science Foundation (NSF) (CNS-1722557, CCF-1718474, OIA-2040667, DGE-1723687 and DGE-1821766) and Intel's gift.
\end{acks}

\bibliographystyle{ACM-Reference-Format}
\bibliography{sample-base}

\appendix









\end{document}